\documentclass[12pt]{article}
\usepackage{amsmath}
\usepackage{graphicx}
\usepackage{amssymb}
\usepackage{amsfonts}
\usepackage{latexsym}
\def\be{\begin{equation}} \def\ee{\end{equation}}
\def\ba{\begin{eqnarray}} \def\ea{\end{eqnarray}} \def\part{\partial}

\begin{document}

\begin{center}
\begin{flushright}\begin{small}    
\end{small} \end{flushright} \vspace{1.5cm}
\huge{Thermodynamics of black plane solution} 
\end{center}

\begin{center}
{\small  Manuel E. Rodrigues $^{(a,f,g)}$}\footnote{E-mail
address: esialg@gmail.com}\,,
{\small  Deborah F. Jardim $^{(b)}$}\footnote{E-mail address: dfjardim@gmail.com}\,,
{\small  St\'{e}phane J. M. Houndjo $^{(c,d)}$}\footnote{E-mail address:
sthoundjo@yahoo.fr} and
{\small Ratbay Myrzakulov$^{(e)}$}\footnote{E-mail address: rmyrzakulov@gmail.com}
\vskip 4mm

(a) \ Universidade Federal do Esp\'{\i}rito Santo \\
Centro de Ci\^{e}ncias
Exatas - Departamento de F\'{\i}sica\\
Av. Fernando Ferrari s/n - Campus de Goiabeiras\\ CEP29075-910 -
Vit\'{o}ria/ES, Brazil \\
(b) \ Universidade Federal dos Vales do Jequitinhonha e Mucuri, ICTM\\
Rua do Cruzeiro, 01, Jardim S\~{a}o Paulo\\
CEP39803-371 - Teofilo Otoni, MG - Brazil\\

(c) \  Departamento de Engenharia e Ci\^{e}ncias Exatas- CEUNES\\
Universidade Federal do Esp\'irito Santo\\
CEP 29933-415 - S\~ao Mateus/ ES, Brazil\\
(d) \ Institut de Math\'ematiques et de Sciences Physiques (IMSP) \\ 01 BP 613 Porto-Novo, B\'enin\\
(e) Eurasian International Center for Theoretical Physics\\
L.N. Gumilyov Eurasian National University, Astana 010008, Kazakhstan\\
(f)\  Faculdade de F\'{\i}sica, Universidade Federal do Par\'{a}, 66075-110, Bel\'{e}m, Par\'{a}, Brazil\\
(g)\  Faculdade de Ci\^{e}ncias Exatas e Tecnologia, Universidade Federal do Par\'{a} - Campus Universit\'{a}rio de Abaetetuba, Rua Manoel de Abreu s/n - Mutir\~{a}o, CEP 68440-000, Abaetetuba, Par\'{a}, Brazil
\vskip 2mm
\end{center}
\vspace{-1cm}
\begin{abstract}
We obtain a new phantom black plane solution in $4$D of the Einstein-Maxwell theory coupled with a cosmological constant. We analyse their basic properties, as well as its causal structure, and obtain the extensive and intensive thermodynamic variables, as well as the specific heat and the first law. Through the specific heat and the so-called geometric methods, we analyse in detail their thermodynamic properties, the extreme and phase transition limits, as well as the local and global stabilities of the system. The normal case is shown with an extreme limit and the phantom one with a phase transition only for null mass, which is physically inaccessible. The systems present local and global stabilities for certain values of the entropy density with respect to the electric charge, for the canonical and grand canonical ensembles.
\end{abstract}
\par 
Pacs numbers: 04.70.-s; 04.20.Jb; 04.70.Dy. 


\section{Introduction}

\hspace{0,5cm} It is well known that a black hole can radiate a black-body radiation when one takes into account the effects of classical gravitational field on quantized matter fields, i.e, a semi-classical analysis of the gravity \cite{hawking1}. So, we can make a study of the thermodynamic system of each new black hole solution. The most common method in the literature is the analysis made through the specific heat of the black hole \cite{davies}, which informs us if the system is thermodynamically interacting, if there exists any case in which the black hole is extreme or it passes across a second order phase transition.
\par
Recently, attention is attached to the methods for analysing the thermodynamic system through the geometry of the so-called thermodynamic space of the equilibrium states. The most common are the methods of Weinhold \cite{weinhold}, Ruppeiner \cite{ruppeiner}, geometrothermodynamics \cite{quevedo} and that of Liu-Lu-Luo-Shao \cite{liu}. These methods also notify if the system possesses thermodynamic interaction and if it undergoes a second order phase transition, in addition to the properties about the stability.
\par
In this work, we desire to make a detailed analysis of the thermodynamic system of a well known class of solutions, with a particularly interesting symmetry, the planar. This class of solutions has been previously obtained for the case of planar and  static symmetry in $4D$, by Cai and Zhang \cite{cai}. This symmetry was then applied to traversable wormholes \cite{lemos}, and later, generalized to topological black holes in \cite{cai2}, and its various applications. We focus our attention to a class of solutions, called phantom \cite{gerard3}, but now with a planar symmetry.   
\par
Before beginning  the analysis of this new class of phantom black holes, we will present briefly our interest in obtaining and studying such exotic solutions. With the discovery of the acceleration of the universe, various observational programs of studying the evolution of our universe were deployed, including the relationship of the magnitude-versus-redshift type supernovae Ia and the spectrum of the anisotropy of the cosmic microwave background. These programs promote an accelerated expansion of our universe, which should be dominated by an exotic fluid and should have a negative pressure. Moreover, these observations show that this fluid can be phantom, i.e, with the contribution of the energy density of dark energy  \cite{hannestad}.
\par
As the interest in obtaining these classes has increased, we also found ourselves wanting to analyse a specific phantom model. We can mention here some recent results in the literature, such as the wormhole solutions and conformal continuation \cite{kirill1}, the black hole solutions of Einstein-Maxwell-Dilaton theory, \cite{gibbons}, the higher-dimensional black holes by Gao and Zhang \cite{gao}, and the higher-dimensional black branes by Grojean et al \cite{grojean}. Analysis were also made in algebraic structures of this type of phantom system, as the case of the algebra generated by metrics depending on two temporal coordinates, with $ D\geq 5$, which provides phantom fields in $4D$, fulfilled by Hull \cite{hull}, and Sigma models by Cl\'{e}ment et al \cite{gerard2}. Here, we will obtain and study the thermodynamic properties of a solution arising from the coupling of Einstein-Hilbert action with a field of spin $1$, which can be Maxwell or anti-Maxwell (phantom), and a cosmological constant, where the spacetime possesses planar symmetry. The idea of using the ruse of negative electric energy density is quit old, Einstein and Rosen being the first to use it \cite{visser}. Recently, through the work of Babichev et al \cite{babichev} and Bronnikov et al \cite{bronnikovPRL}, we have seen a keen interest in phantom solutions \cite{phantom}. 
\par
The paper is organized as follows. In Section \ref{sec2}, we present a new phantom black plane solution. The causal structure of the solutions are studied and the thermodynamic variables are obtained. The first law of thermodynamics is established and the specific heat is calculated. In Section \ref{sec3}, we minutely study the thermodynamics of normal and phantom solutions, using the analysis through the specific heat, subsection \ref{subsec3.1}, and through the geometric methods of Weinhold, subsection \ref{subsec3.2}, the geometrothermodynamics, subsection \ref{subsec3.3}, and that of  Liu-Lu-Luo-Shao, subsection \ref{subsec3.4}. We finish the section with the study of local and global stabilities in subsection \ref{subsec3.5}. The conclusion is presented in  Section \ref{sec4}.  


\section{ The field equations and the black holes solutions}\label{sec2}
\hspace{0,5 cm} The action of the theory is given by: 
\be
S=\int d^{4}x\sqrt{-g}\left[  \mathcal{R}+\eta F^{\mu\nu}F_{\mu\nu}+2\Lambda\right]  \label{action1}\; ,
\end{equation}
where the first term is that of Einstein-Hilbert, the second is the coupling of  (anti)Maxwell field $F_{\mu\nu}=\partial_{\mu}A_{\mu}-\partial_{\nu}A_{\mu}$ with the gravitation, and the third is the cosmological constant. Making the functional variation of the action (\ref{action1}) with respect to the field  $A_{\mu}$ and the inverse of the metric, $g^{\mu\nu}$, using  $R=-4\Lambda$, we get the following equations of motion 
\begin{eqnarray}
\nabla_\mu\left[ F^{\mu\alpha}\right]&=&0\label{em1}\; ,\\
R_{\mu\nu}&=&2\eta\left(\frac {1}{4}g_{\mu\nu}F^{2}
-F_{\mu}^{\;\;\sigma}F_{\nu\sigma}\right)-\Lambda g_{\mu\nu}\label{em2}\, .
\end{eqnarray}

Let us write the static and plane symmetric line element as
\begin{equation}
dS^{2}=A(r)dt^{2}-B(r)dr^{2}-C(r)(dx^2+dy^2)
\label{metric}\; ,
\end{equation}
with  $r=|z|$. We will also assume that the Maxwell field is purely electric and only depends on  $r$. With (\ref{metric}), one can integrate (\ref{em1}) and obtain 
\begin{equation}
F ^{10}(r)=\frac{q}{C\sqrt{AB}} \qquad
(F ^{2}=-2\frac{q^{2}}{C^2}) \label{1}\; ,
\end{equation}
with $q$ a real integration constant. Substituting  (\ref{1}) into the equations of motion (\ref{em2}), we
obtain the equations 
\begin{eqnarray}
\frac{A^{\prime\prime}}{A}-\frac{1}{2}\left(\frac{A^{\prime}}{A}\right)^2-\frac{A^{\prime}B^{\prime}}{2AB}+\frac{A^{\prime}C^{\prime}}{AC}=2B\left(\eta\frac{q^2}{C^2}-\Lambda\right)\label{em3} ,\\
\frac{A^{\prime\prime}}{A}-\frac{1}{2}\left(\frac{A^{\prime}}{A}\right)^2-\frac{A^{\prime}B^{\prime}}{2AB}+2\frac{C^{\prime\prime}}{C}-\frac{B^{\prime}C^{\prime}}{BC}-\left(\frac{C^{\prime}}{C}\right)=2B\left(\eta\frac{q^2}{C^2}-\Lambda\right)\label{em4} ,\\
-\frac{A^{\prime}C^{\prime}}{2AC}-\frac{C^{\prime\prime}}{C}+\frac{B^{\prime}C^{\prime}}{2BC}=2B\left(\eta\frac{q^2}{C^2}+\Lambda\right)\label{em5} ,
\end{eqnarray}
where the ``prime"  denotes the derivative with respect to  $r$. Choosing the coordinates such that 
\begin{equation}
A(r)=B^{-1}(r)\,,\,C(r)=\alpha^2 r^2\,
\end{equation}
with  $\Lambda=-3\alpha^2$, the solution of the equations of motion (\ref{em3})-(\ref{em5}) is given by 
\begin{eqnarray}
\left\{\begin{array}{ll}
dS^2=A(r)dt^2-A^{-1}(r)dr^2-C(r)(dx^2+dy^2)\,,\\
F=-\frac{q^2}{C(r)}dr\wedge dt\,,\,A(r)=\alpha^2 r^2-\frac{m}{r}+\eta\frac{q^2}{\alpha^4r^2}\,,\,C(r)=\alpha^2r^2\,,
\end{array}\right.
\end{eqnarray}
where  $m$ is the mass and $q$ the electric charge of the (phantom) black plane. This is the same solution as that of \cite{cai}, for $\eta=1$, and phantom black plane solution for $\eta=-1$, obtained for the first time here. 
\par
We can rewrite the solution in terms of the densities of mass $M$ and electric charge  $Q$, as calculated in \cite{cai}, yielding 
\begin{eqnarray}\label{plane}
\left\{\begin{array}{ll}
dS^2=A(r)dt^2-A^{-1}(r)dr^2-C(r)(dx^2+dy^2)
\,,\\
F=-\frac{2\pi Q}{C(r)}dr\wedge dt\,,\,A(r)=\alpha^2 r^2-\frac{4\pi M}{\alpha^2r}+\eta\frac{4\pi^2 Q^2}{\alpha^4r^2}\,,\,C(r)=\alpha^2r^2\,.
\end{array}\right.
\end{eqnarray}

\begin{figure}[h]
\begin{center}
\includegraphics[height=6cm,
width=10cm]{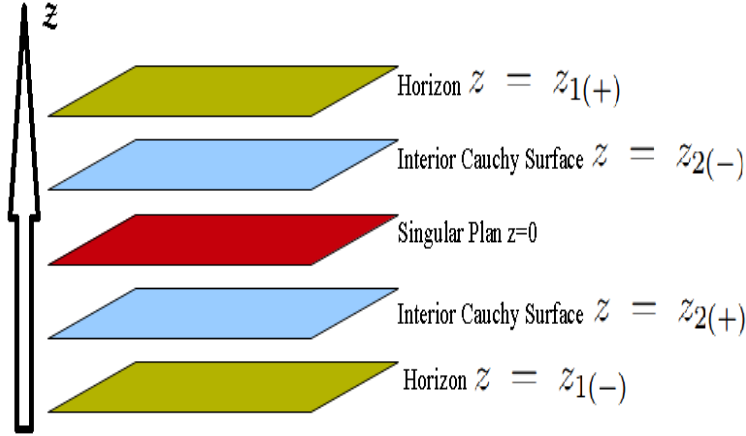}
\end{center}
\caption{\scriptsize{Structure of spacetime in $z$ direction, for the phantom solution (\ref{plane}).}} \label{fig1}
\end{figure}

One can calculate the horizon of this solution, vanishing $A(r)$, obtaining 
\begin{eqnarray}
\alpha^2 r^2-\frac{4\pi M}{\alpha^2r}+\eta\frac{4\pi^2 Q^2}{\alpha^4r^2}=0\,.\label{horizon}
\end{eqnarray}
This solution possesses two complex and two real roots. The real roots are given by  
\begin{eqnarray}
r_{\pm}&=&\frac{1}{2}\left[\sqrt{2k}\pm\sqrt{\frac{8\pi M}{\alpha^4 \sqrt{2k}}-2k}\right]\label{r1}\,,\\
k&=&\sqrt[3]{\left(\frac{\pi M}{\alpha^4}\right)^2+\sqrt{\left(\frac{\pi M}{\alpha^4}\right)^4-\eta\left(\frac{4\pi^2 Q^2}{3\alpha^6}\right)^3}}\nonumber\\
&&+\sqrt[3]{\left(\frac{\pi M}{\alpha^4}\right)^2-\sqrt{\left(\frac{\pi M}{\alpha^4}\right)^4-\eta\left(\frac{4\pi^2 Q^2}{3\alpha^6}\right)^3}}\,.
\end{eqnarray}
For the normal solution, $\eta=1$, one has $0<r_{-}<r_{+}$, and for $\eta=-1$, the corresponding is $r_{-}<0<r_{+}$, with $r_{+}>|r_{-}|$. We observe that  in the  phantom solution, $r_-$ is in the negative part, but here something happens that we do not have in the spherical symmetry, because as $r_{\pm}=|z_{1,2}|$, one gets $z_{1(\pm)}=\pm r_{+}$  and $z_{2(\pm)}=\pm r_{-}$. As $r_{-}<0$, one gets $z_{1(-)}<z_{2(+)}<0<z_{2(-)}<z_{1(+)}$. Then, the singular plan $z=r_s=z_s=0$ is covered by the plans $z=z_{1(-)},z=z_{2(+)},z=z_{2(-)}$ and $z=z_{1(+)}$ (see Figure \ref{fig1}). In the case of spherical symmetry, the internal horizon $r_{-}$ could not be achieved, for a solution of non-degenerate horizon. Hence, here we have a drastic change in the causal structure of the phantom black plane solution, whose singularity is covered by two horizons in the positive part of $z$. This could not occur in the phantom solutions with spherical symmetry, where just one horizon covered the singularity. However, another unusual event happens, where we get two horizons but with the property of non existence of extreme case, i.e, these horizons can never be equal, when we consider only real values.
\par
The curvature scalar of the metric (\ref{metric}) is given by
\begin{eqnarray}
R=\frac{2C'' }{BC}-\frac{{\left(C'\right) }^{2}}{2B{C}^{2}}-\frac{B'C'}{{B}^{2}C}+\frac{A'C'}{ABC}-\frac{A'B' }{2A{B}^{2}}+\frac{A''}{AB}-\frac{\left(A'\right) ^{2}}{2{A}^{2}B}.
\end{eqnarray}
The scalar of Kretschmann is given by
\begin{eqnarray}
&&K=R^{\mu\nu\gamma\delta}
R_{\mu\nu\gamma\delta}={C}^{2}\left(C''\right)^{2}+{B}^{2}
\left(C''\right)^{2}-C\left(C'\right)^{2}C'' -\frac{B^{2}\left(C'\right)^{2}C'' }{C}-\frac{B'C'C''{C}^{2} }{B}\nonumber\\
&&-BB'C'C''+\frac{{C}^{2}\left(C'\right)^{4}}{4{B}^{2}}+\frac{{B}^{2}\left(C'\right)^{4}}{4{C}^{2}}+\frac{{C' }^{4}}{4}+\frac{B'C\left(C'\right)^{3}}{2B}+\frac{BB'\left(C'\right)^{3}}{2C}\nonumber\\
&&+\frac{\left(B'\right) ^{2}{C}^{2}\left(C'\right) ^{2}}{4{B}^{2}}+\frac{\left( A'\right) ^{2}{C}^{2}\left(C'\right)^{2}}{4{B}^{2}}+\frac{\left(B'\right)^{2}
\left(C'\right)^{2}}{4}+\frac{{A}^{2}\left(A'\right)^{2}
\left(C'\right)^{2}}{4{B}^{2}}+\frac{{A}^{2}\left(A'\right)^{2}
\left(B'\right)^{2}}{8{B}^{2}}\nonumber
\end{eqnarray}
\begin{eqnarray}
&&+\frac{\left(A'\right)^{2}\left(B'\right) ^{2}}{8}-\frac{A' A''BB'}{2}+\frac{\left(A'\right)^{3}BB' }{4A}-\frac{{A}^{2}A'A''B' }{2B}+\frac{A\left(A'\right)^{3}B'}{4B}+\frac{\left(A''\right)^{2}{B}^{2}}{2}\nonumber\\
&&-\frac{\left(A'\right)^{2}A''{B}^{2}}{2A}+\frac{\left(A'\right)^{4}{B}^{2}}{8{A}^{2}}+\frac{{A}^{2}\left(A''\right)^{2}}{2}-\frac{A\left(A'\right)^{2}A''}{2}+\frac{\left(A'\right)^{4}}{8}.
\end{eqnarray}
By substituting $A(r)=B^{-1}(r)$ and  $C(r)$ in  (\ref{plane}), the curvature scalar ($R=12\alpha^2$) and that of Kretschmann are finite throughout the space-time, except in the singular plane $r_s=z=0$.\par
In order to construct the Penrose diagram of this solution, we define several new coordinates for getting a description (non-singular on the horizons) of this space-time of type Kruskal. So, the Eddington-Finkelstein coordinates are gives by
\begin{eqnarray}
u=t+r^*,v=t-r^*\,,
\end{eqnarray}  
where the tortoise coordinate is give by
\begin{eqnarray}
&&r^*=\int A^{-1}(r)dr=\frac{1}{\alpha^2}\Big\{ \frac{1}{r_+-r_-}\ln\left|\frac{r-r_+}{r-r_-}\right|-\frac{(r_++r_-)^2+r_+^2}{(r_+-r_-)[(r_+^2+r_-)^2+2r_+^2]}\times\nonumber\\
&&\times \ln|r-r_+|+\frac{(r_++r_-)^4+2(r_+^2+r_-^2)^2+2r_+r_-(r_+^2+r_-^2)}{[(r_++r_-)^2+2r_+^2][(r_++r_-)^2+2r_-^2]\sqrt{(r_++r_-)^2+2(r_+^2+r_-^2)}}\nonumber\\
&&\times \arctan\left(\frac{2r+r_++r_-}{\sqrt{(r_++r_-)^2+2(r_+^2+r_-^2)}}\right)+\frac{(r_++r_-)^2+r_-^2}{(r_+-r_-)[(r_+^2+r_-)^2+2r_-^2]} \ln|r-r_-|\nonumber\\
&&\frac{(r_++r_-)^3}{4(r_++r_-)^4+2(r_+^2+r_-^2)^2}\ln|r^2+(r_+-r_-)r+(r_++r_-)^2-r_+r_-|\Big\}.
\end{eqnarray} 
With these coordinates, we can rewrite the line element (\ref{plane}) as
\begin{eqnarray}
dS^2=A(r)du^2+2dudv-C(r)\left(dx^2+dy^2\right)\,.\label{EF}
\end{eqnarray}
Also defining the coordinates of type Kruskal 
\begin{eqnarray}
&&U=\arctan\left\{\mp k_0\exp\left[-\frac{\alpha^2}{2}(r_+-r_-)[2+(1+k_1)^2]v\right]\right\}\label{U}\,,\\
&&V=\arctan\left\{\pm k_0\exp\left[\frac{\alpha^2}{2}(r_+-r_-)[2+(1+k_1)^2]u\right]\right\}\label{V}\,,\\
&&k_1=\frac{r_+}{r_-}\,,\,k_0=\frac{r_-^{k_1}}{\sqrt{r_+}}(r_+^2+r_+r_-+r_-^2)-\left(\frac{1-k_1}{2}\right)\frac{(1+k_1)^3[2+(1+k_1)^2]}{4(1+k_1)^4+2(1+k_1^2)^2}\times\nonumber\\
&&\times\exp\Big\{-\left(\frac{1-k_1}{2}\right)\frac{(1+k_1)^4+2(1+k_1^2)(1+k_1+k_1^2)}{[(1+k_1)^2+2k_1^2]\sqrt{(1+k_1)^2+
2(1+k_1^2)^2}}\times
\end{eqnarray}
\begin{eqnarray}
&&\times\arctan\left(\frac{1+k_1}{\sqrt{(1+k_1)^2+2(1+k_1^2)}}\right)\Big\}
\end{eqnarray}
we can rewrite (\ref{EF}) as
\begin{eqnarray}
dS^2=\Omega(U,V)dUdV-C(r)\left(dx^2+dy^2\right)\,.\label{K}
\end{eqnarray}
\par
\begin{figure}[h]
\begin{center}
\begin{tabular}{rl}
\includegraphics[height=6cm,width=10cm]{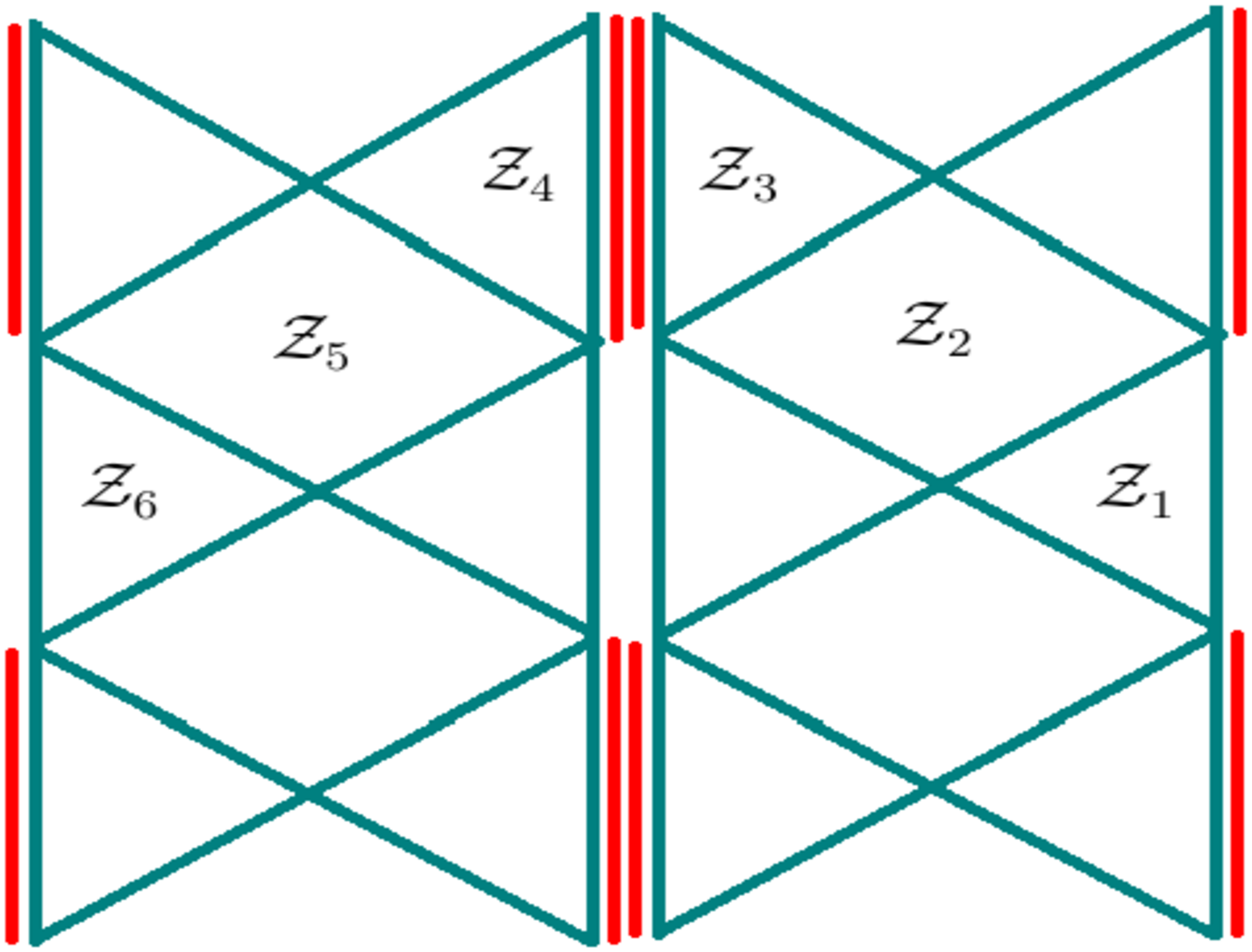}
\end{tabular}
\caption{\scriptsize{Penrose diagram for phantom black plane solution (\ref{plane}).}} \label{fig2}
\end{center}
\end{figure}

With the use of these coordinates we can construct the causal structure of this solution, which is very similar to the Reissner-Nordstrom-AdS one (see Figure \ref{fig2}).
\par
We can see in Figure \ref{fig2} that if we think to follow the decreasing $z$, starting from positive infinity, we have the region $\mathcal{Z}_1$ ($z_{1(+)}<z<+\infty$), passing by the first horizon at $z=z_{1(+)}$, for the second region $\mathcal{Z}_2$ ($z_{2(-)}<z<z_{1(+)}$). After we passed the second horizon at $z=z_{2(-)}$, for the third region $\mathcal{Z}_3$ ($0\leq z <z_{2(-)}$).
\par
After arriving at the singular plane at $z=0$\footnote{This singularity is timelike and can be avoided, depending on the particle energy applied in its path.}. These regions $z\geq 0$ are causally disconnected from those for which $z\leq 0$. Regions from $\mathcal{Z}_4$ to $\mathcal{Z}_6$ are the exact reflection (symmetrical values of positive $z$) for positive values of $z$. So, we can think alike to follow a direction of creasing values of $z$, beginning at negative infinity. Thus, we perform the reflected route, and spent from $\mathcal{Z}_6$ ($-\infty<z<z_{1(-)}$) to $ \mathcal{Z}_5$ ($z_{1(-)} <z <z_{2(+)}$), and then, to the region $\mathcal{Z}_4$ ($ z_{2(+)}<z\leq 0$), reaching the singular plane at  $z=0$.    
\par
Now, we are interested in the geometrical analysis representing semi-classical gravitational effects of the black hole solutions as mentioned before. By semi-classical we mean quantize the called matter fields, while the background gravitational field is treated classically. Therefore, we will work with the semi-classic thermodynamics of black holes, studied first by Hawking \cite{hawking1}, and further developed by many other authors \cite{davies2}.
\par
There are several techniques to derive the Hawking temperature law. For example we can mention the Bogoliubov coefficients \cite{ford} and the energy-momentum tensor methods \cite{davies,davies2}, the euclidianization of the metric \cite{hawking2}, the transmission and reflection coefficients \cite{gerard4,kanti}, the analysis of the anomaly term \cite{robinson}, and the black hole superficial gravity \cite{jacobson}. Since all these methods have been proved to be equivalent \cite{glauber}, then we opt, without loss of generality, to calculate the Hawking temperature by the superficial gravity method.
\par
The surface gravity of a black plane is given by \cite{cai}:
\begin{equation}
\kappa =\left[  \frac{g_{00}^{\prime}}{2\sqrt{-g_{00}g_{11}}}\right]_{r\;=\;r_{+}}\, ,\label{10}
\end{equation}
where $r_{+}\;$ is the event horizon radius, and the Hawking temperature is related with the surface  gravity through the relationship \cite{hawking1,jacobson}
\begin{equation}
T=\frac{\kappa}{2\pi}\, .\label{11}
\end{equation}
\par
Then, for the black plane solution (\ref{plane}), we get the surface gravity (\ref{10}) as 
\begin{equation}
 \kappa =\alpha^2r_{+}+\frac{2\pi M}{\alpha^2 r_{+}^2}-\eta\frac{4\pi^2 Q^2}{\alpha^4 r_{+}^3}\,  ,\label{12}
\end{equation}
and the Hawking temperature (\ref{11}) in this case is :
\begin{equation}
T=\frac{1}{2\pi}\left[\alpha^2r_{+}+\frac{2\pi M}{\alpha^2 r_{+}^2}-\eta\frac{4\pi^2 Q^2}{\alpha^4 r_{+}^3}\right]\, .\label{13}
\end{equation}
We define the entropy per unit of area of the black plane as two times the quarter of the horizon area
\begin{eqnarray}
S=2\times\frac{1}{4}A=\frac{\alpha^2r_{+}^2}{2}\,,\label{s}
\end{eqnarray}
where the factor $2$ is due to the contribution of two planes  $z=\pm r_{+}$.
\par
From (\ref{plane}), we can calculate the electric potential scalar at the horizon 
\begin{eqnarray}
A_{0}=\int\limits^{r}_{+\infty}F_{10}(r^{\prime})dr^{\prime}\Big|_{r=r_{+}}=\frac{2\pi Q}{\alpha^2 r_{+}}\,.\label{14}
\end{eqnarray}
\par
Let us check the first law for the solution (\ref{plane}). Taking the differential of the mass, isolated from (\ref{horizon}), of the electric charge and of the entropy (\ref{s}), we get
\begin{eqnarray}
dM=\left(\frac{3\alpha^4 r_{+}^2}{4\pi}-\eta\frac{\pi Q^2}{\alpha^2 r_{+}^2}\right)dr_{+}+\eta\frac{2\pi Q}{\alpha^2 r_{+}}dQ\label{dmass}\,,\,dS=\alpha^2 r_{+}dr_{+}\label{ds}\,,
\end{eqnarray}
which satisfies the first law of thermodynamics  
\begin{eqnarray}
dM=TdS+\eta A_{0}dq\,.\label{pl}
\end{eqnarray}
Note that we introduced a compensating sign $\eta$ in (\ref{pl}) due to the contribution of the negative energy density, in the phantom case, the field of spin $1$, $F_{\mu\nu}$, which provides  a work with an inverted sign in the first law. 
\par
As we need to study the thermodynamic system through the geometric methods, we must first write the mass in terms of the entropy and the electric charge. We can do this by isolating the mass in (\ref{horizon}) and then replace $r_{+}$ in terms of the entropy\footnote{We take $r_{+}=\sqrt{2S}/\alpha$, with the sign of  $\alpha>0$. The negative sign of  $\alpha$ can be considered taking $r_{+}=-\sqrt{2S}/\alpha$.}, with the use of (\ref{s}), which yields 
\begin{eqnarray}
M(S,Q)=\frac{\alpha^2S^2+\eta\pi^2 Q^2}{\pi\alpha\sqrt{2S}}\label{mass}\,,
\end{eqnarray}
where we have the conditions $Q^2\leq(3\alpha^6/4\pi^2)(\pi M/\alpha^4)^{4/3}$ for  $\eta=1$ \cite{cai} (real horizon in  (\ref{r1})) and  $Q^2\leq(\alpha^2S^2/\pi^2)$ for $\eta=-1$. We also write the temperature and the electric potential in terms of the entropy and the electric charge. Taking (\ref{13}) and (\ref{14}), for $r_{+}$ in terms of the entropy, we get
\begin{eqnarray}
T(S,Q)&=&\frac{3\alpha^2S^2-\eta\pi^2 Q^2}{\pi\alpha (\sqrt{2S})^{3}}\,,\,A_0=\frac{2\pi Q}{\alpha\sqrt{2S}}\label{t1}.
\end{eqnarray}
\par
We can then calculate the specific heat by the expression 
\begin{equation}
C_{Q}=\left(\frac{\partial M}{\partial T}\right)_{Q}=\left(\frac{\partial M}{\partial S}\right)_{Q}\Big/\left(\frac{\partial^2 M}{\partial S^2}\right)_{Q}=\frac{2S}{3}\frac{(3\alpha^2S^2-\eta\pi^2 Q^2)}{(\alpha^2S^2+\eta\pi^2 Q^2)}\label{cqm}\;.
\end{equation}
\par 
We now have in hand  the basic requirements to begin our analysis of the thermodynamic system of these solutions. In the next section we will study the specific heat (\ref{cqm})  and through the four geometric methods, the thermodynamic properties of these planar solutions.  
\section{ Thermodynamics of black plane}\label{sec3}
\hspace{0,5cm} In this section we will study in detail the thermodynamic properties of the planar solutions (\ref{plane}), both for normal and phantom cases. Through the specific heat and the curvature scalar of the thermodynamic spaces of the equilibrium states, we will examine whether there is an extreme case (only by the usual method), phase transition and finally, the local and global stabilities of the thermodynamic system.
\subsection{ Analysis of specific heat}\label{subsec3.1}
\hspace{0,5cm} Historically, the study of specific heat for revealing the thermodynamic properties was the first to be used \cite{davies} and has been called of usual method.
\par
Here, we have the expression of the specific heat (\ref{cqm}), which, equating to zero, reveals the value of the entropy for which the solution is  extreme, i.e, for $S=S_{e}=\pi Q\sqrt{\eta}/\alpha\sqrt{3}$, which is real only for  $\eta=1$. Therefore, there does not exist an extreme case for the  phantom solution with $\eta=-1$, as we had seen in its causal structure.
\par
Similarly, we can find the value of the entropy for which the system undergoes a second order phase transition, i.e, when the specific heat diverges. In this case the specific heat (\ref{cqm}) diverges for $S=S_{t}=-i\sqrt{\eta}\pi Q/\alpha$, which shows that the normal case $ \eta=1$ has no phase transition, while the phantom case possesses a phase transition in  $S=S_{t}$. Note that this  case is the specific value where the  mass (\ref{mass}) vanishes. So, here, we have a mathematical chance of the system going from a locally stable phase ($C_{Q}> 0$ and positive mass), for an unstable phase, with $C_{Q}<0$ and negative mass (\ref{mass}). The phase transition of second order is not physically possible because the energy of the phantom black plane should be reduced continuously such that it passes from the positive values to zero, and even reaching negative values. This will be well examined in the stability study of the system.
\par
\begin{figure}[h]
\centering
\begin{tabular}{rl}
\includegraphics[height=4cm,width=10cm]{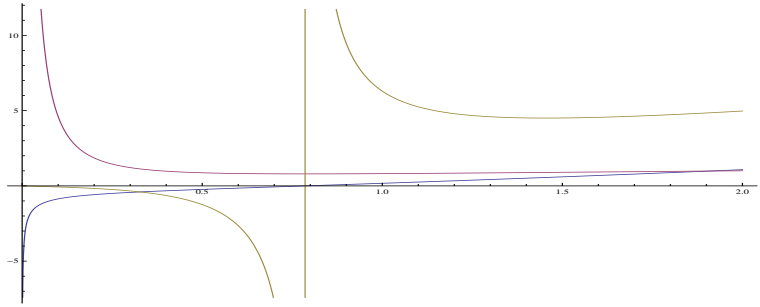}
\end{tabular}
\caption{\scriptsize{The mass (\ref{mass}) (blue), the temperature (\ref{t1}) (purple) and specific heat (\ref{cqm}) (green) for $Q=0.5,\alpha=2,\eta=-1$.  The phase transition point is given by $S_t=0.785398$.}} \label{fig3}
\end{figure}

We plot the evolution of the specific heat (\ref{cqm}) for a specific choice of the parameters, as shown in Figure \ref{fig3}.
\par
We will take the results of the study of specific heat as the basis for comparing with a geometric analysis of the thermodynamic system, through the four most popular methods in the literature. All these methods have in common the definition of a metric for the thermodynamic space of the equilibrium states, where the calculation of the curvature scalar of this metric reveals the existence or not of thermodynamic interaction, phase transition points, among other thermodynamic properties. Let us calculate this object with the aid of a mathematical software.
\par
In the next subsection we will analyse the thermodynamic system through the method of Weinhold.

\subsection{ The Weinhold method}\label{subsec3.2}
\hspace{0,5cm} Historically, Weinhold was one of the first to formulate a  geometric description applicable to a thermodynamic system. The method of Weinhold \cite{weinhold}, as it is known, aims to define a metric for the thermodynamic space of the equilibrium states, through the mass (\ref{mass}) as thermodynamic potential. The metric constructed in this way provides a curvature scalar $R_W$, which, for this method can be interpreted as a function of extensive variables that shows the points of phase transition, when there exists, where the thermodynamic system goes by. Then, we define the metric of Weinhold as being
\begin{eqnarray}
dl^{2}_{W}&=&\frac{\partial^2 M}{\partial S^2}dS^2+2\frac{\partial^2 M}{\partial S\partial Q}dSdQ+\frac{\partial^2 M}{\partial Q^2}dQ^2\nonumber\\
&=&\frac{3(\alpha^2 S^2+\eta\pi^2 Q^2)}{4\sqrt{2}\pi\alpha S^{5/2}}dS^2-\frac{2\eta\pi Q}{\sqrt{2}\alpha S^{3/2}}dSdQ+\frac{\eta\sqrt{2}\pi}{\alpha\sqrt{S}}dQ^2\label{mw}\;.
\end{eqnarray}
Here we see that the curvature scalar $R_{W}$ of this metric is identically zero, which prevents us of doing an analysis of the phase transition of the thermodynamic system. This result does not agree with the study of the specific heat. In the next subsection we will study the thermodynamics through the method of geometrothermodynamics.


\subsection{ The Geometrothermodynamics method}\label{subsec3.3}
\hspace{0,5cm}The Geometrothermodynamics (GTD) \cite{quevedo} makes use of differential geometry as a tool to represent the thermodynamics of physical systems. Let us consider the $(2n+1)$-dimensional space $\mathbb{T}$, whose coordinates are represented by the thermodynamic potential $\Phi$, the extensive variable $E^{a}$ and the intensive variables $I^{a}$, where $a=1,...,n$. If the space $\mathbb{T}$ has a non degenerate metric $G_{AB}(Z^{C})$, where $Z^{C}=\{ \Phi , E^{a} , I^{a}\}$, and the so called Gibbs 1-form  $\Theta=d\Phi -\delta_{ab}I^{a}dE^{b}$, with $\delta_{ab}$ the delta Kronecker; then, the structure $\left(\mathbb{T} , \Theta, G\right)$ is said to be a contact riemannian manifold if $\Theta\wedge \left(d\Theta\right)^{n}\neq 0$ is satisfied \cite{hermann}. The space $\mathbb{T}$ is known as the thermodynamic phase space. We can define a $n$-dimensional subspace $\mathbb{E}\subset \mathbb{T}$, with extensive coordinates $E^{a}$, by the map $\varphi :\mathbb{E}\rightarrow\mathbb{T}$, with $\Phi\equiv \Phi (E^{a})$, such that $\varphi^{*}(\Theta)\equiv 0$. We call the space $\mathbb{E}$ the thermodynamic space of the equilibrium states.
\par
We can then define the metric of the thermodynamic space of the equilibrium states $\mathbb{E}$, through the derivation of the thermodynamic potential and its extensive variables as \cite{zui}
\begin{align}
dl^{2}_{G(\Phi)} & =\left( E^{c}\frac{\partial\Phi}{\partial E^{c}}\right) \left( \eta_{ad}\delta^{di}\frac{\partial^{2}\Phi}{\partial E^{i}E^{b}}\right)dE^{a}dE^{b}\;,\label{me}
\end{align}
which, by definition, is invariant under Legendre transformations.
Through the metric (\ref{me}), we can calculate the curvature scalar of the space $\mathbb{E}$, which informs if the system passes by a phase transition, when the scalar diverges for some value of extensive coordinates. If the scalar is not zero, the system possesses  thermodynamic interaction, i.e, the Hawking temperature is non null.
\par
Here, we will do the calculation of the metric of $\mathbb{E}$, using the mass (\ref{mass}) as the thermodynamic potential, which provides
\begin{eqnarray}
dl^{2}_{G(M)}&=&-\frac{9(\alpha^2 S^2+\eta\pi^2 Q^2)^2}{16\alpha^2\pi^2 S^3}dS^2+\frac{3\eta(\alpha^2 S^2+\eta\pi^2 Q^2)}{2\alpha^2 S}dQ^2\label{mem}\;.
\end{eqnarray}
The curvature scalar of this metric is given by
\begin{eqnarray}
R_{G}=\frac{8\alpha^2\pi^2 S^3}{9}\frac{(-5\alpha^2 S^2+7\eta\pi^2 Q^2)}{(\alpha^2 S^2+\eta\pi^2 Q^2)^4}\label{rg}\,.
\end{eqnarray}
We get the value for which the scalar (\ref{rg}) diverges, which is given by $S_{t}=-i\sqrt{\eta}\pi Q/\alpha $, in agreement with the value obtained through the specific heat (\ref{cqm}). This result is consistent with the specific heat, where we have found that the normal case has no phase transition and in the phantom case has one point of second order phase transition in $S=S_t$. 
\par
In the next subsection we will see the analysis made by the geometric method of Liu-Lu-Luo-Shao.


\subsection{ The Liu-Lu-Luo-Shao method}\label{subsec3.4}
\hspace{0,5cm} The geometric method of the analysis of the more recent thermodynamic system is that of Liu-Lu-Luo-Shao \cite{liu}, which defines a metric in the thermodynamic space of the equilibrium states, based on the Hessian matrix of several free energy, the Helmholtz's one in our case, and which can be written as follows
\begin{eqnarray}
dl^{2}_{LLLS(F)}&=&-dTdS+\eta dA_{0}dq=-\frac{\partial T}{\partial S}dS^2+\left(\eta\frac{\partial A_{0}}{\partial S}-\frac{\partial T}{\partial q}\right)dSdq+\eta\frac{\partial A_{0}}{\partial q}dq^2\nonumber\\
&=&-\frac{3(\alpha^2 S^2+\eta\pi^2 Q^2)}{4\sqrt{2}\alpha\pi S^{5/2}}dS^2+\eta\frac{\sqrt{2}\pi}{\alpha\sqrt{S}}dQ^2\label{mllls}\;.
\end{eqnarray}

The curvature scalar of this metric is given by  
\begin{eqnarray}
R_{LLLS}=-\frac{\sqrt{2}\alpha^3\pi S^{5/2}}{3\left(\alpha^2 S^2+\eta\pi^2 Q^2\right)^2}\label{rllls}\,.
\end{eqnarray}
Then, the analysis by this method shows that the normal case does not possess  phase transition and  the phantom case possess a transition phase at $S=S_t=-i\sqrt{\eta}\pi Q/\alpha$, which is in agreement with the specific heat.
\par
In the next subsection we will study the local and global stabilities of the black plane solutions.


\subsection{ The local and global stability}\label{subsec3.5}
\hspace{0,5cm} Let us now study the local and global stabilities of these solutions. Through the specific heat (\ref{cqm})\footnote{we can do $C_q=(4S/3)[T(S,Q)/M(S,Q)]$.} and the temperature (\ref{t1}), one can see that in the normal case, $\eta=1$, the system is locally stable for $3\alpha^2 S^2>\pi^2 Q^2$, with $C_{q},T>0$, and unstable for the other values. In the phantom case, $\eta=-1$, the system presents a local stability for $\alpha^2 S^2>\pi^2 Q^2$, with $C_{q},T,M>0$ (see Figure \ref{fig3}). 
\par
Defining the Gibbs's potential 
\begin{eqnarray}
G=M-TS-\eta A_0 Q=-\left(\frac{\alpha^2 S^2+\eta\pi^2 Q^2}{2\pi\alpha\sqrt{2S}}\right)=-\frac{M}{2}\label{gibbs1}\,,
\end{eqnarray}
we get that in the normal case, in the grand canonical ensemble, the system is globally stable for any values of $S$ and  $Q$, with $G<0,\forall  S, Q $. But in the phantom case, the system is globally stable only if $\alpha^2 S^2>\pi^2 Q^2$, which agrees with the local stability of the specific heat.   
\par
Here it is clear that both the specific heat and the Gibbs potential are closely linked to the sign of the mass (\ref{mass}). We have already seen from the specific heat that the mass value, zero, is precisely the point of phase transition of the phantom case. Here, it is also clear from the Gibbs potential that, passing to the negative values of the energy (mass), the system is unstable, not only locally, but also globally. This shows that the system can not move to that physically impossible stage. The explanation is that, when the system loses its energy, approaching zero, this should be treated by a more elaborated quantization, and not a simple semi-classical analysis, as we see here. Thus, we can conclude here that the phase transition presented by the phantom case, is nothing more than a purely mathematical transition, showing a divergence in the specific heat, but which is physically inaccessible to the states of the thermodynamic system. 
\par
In the canonical ensemble, we can define the  Helmholtz free energy as
\begin{eqnarray}
F=M-TS=-\left(\frac{\alpha^2 S^2-3\eta \pi^2 Q^2}{2\pi\alpha\sqrt{2S}}\right)\label{Hel}\,,
\end{eqnarray}
which yields a globally stable system ($F<0$), for the normal case, when $\alpha^2 S^2>3\pi^2 Q^2$, and for the phantom case $F<0,\forall S,Q$.
 
\section{ Conclusion}\label{sec4}


\hspace{0,5cm} We obtained a new phantom black plane solutions in (\ref{plane}). We analysed their basic geometric properties, the causal structure, obtaining the thermodynamic variables, temperature (\ref{13}), entropy density (\ref{s}) and the electric potential (\ref{14}). We established the first law of thermodynamics in (\ref{pl}) and calculated the specific heat (\ref{cqm}).
\par
We analysed the thermodynamic system through the study of the specific heat and the geometric methods called Weinhold, the geometrothermodynamics and that of Liu-Lu-Luo-Shao. In the Weinhold's case, the space metric is not invariant under Legendre transformations, and thus cannot reconcile a good thermodynamic analysis, therefore, in general, this method cannot agree with that of specific heat. By the use of the geometrothermodynamics and the method of Liu-Luo-Shao, we obtain the same results as in the case of specific heat, which shows that these two geometric methods agree with the usual one. 
\par
The summarized results are that the normal case possesses an extreme limit for $S=S_{e}=\pi Q\sqrt{\eta}/\alpha\sqrt{3}$, and the phantom case presents a phase transition point in  $S=S_{t}=-i\sqrt{\eta}\pi Q/\alpha$, which represents a solution with mass  (\ref{mass}) identically null. The interpretation of massless solutions has been presented in \cite{gerard4}, but without any conclusion about its thermodynamics.
\par 
The normal case presents locally stable thermodynamic system, for $3\alpha^2 S^2>\pi^2 Q^2$, and globally stable, in grand canonical ensemble, when $G<0,\forall S, Q$, and in  canonical ensemble for $\alpha^2 S^2>3\pi^2 Q^2$. On the other hand, the phantom  case is locally stable when  $\alpha^2 S^2>\pi^2 Q^2$, and globally stable, in grand canonical ensemble, when $\alpha^2 S^2>\pi^2 Q^2$, and in canonical ensemble, when $F<0,\forall S,Q$.
\par
We conclude with the most important result here, which is the demonstration that normal and phantom cases have no physical phase transition, and that the normal case is an extreme case but not the phantom one.

\vspace{0,2cm}

{\bf Acknowledgement}:  M. E. Rodrigues thanks a lot UFES and PPGF of the UFPA for the hospitality during the elaboration of this work and also CNPq for financial support. S. J. M. Houndjo thanks CNPq/FAPES for financial support.



\begin{thebibliography}{10}
\bibitem{hawking1} S. Hawking, Commun. Math. Phys. {\bf 43}, 199 (1975).
\bibitem{davies} P. C. W. Davies, Proc.Roy.Soc.Lond. A {\bf 353}: 499-521 (1977).
\bibitem{weinhold} F. Weinhold, J. Chem. Phys. {\bf 63}, 2479, 2484, 2488, 2496 (1975); {\bf 65}, 559 (1976).
\bibitem{ruppeiner}G. Ruppeiner, Phys. Rev. A {\bf 20}, 1608 (1979); Rev. Mod. Phys. {\bf 67}, 605 (1995); {\bf 68}, 313 (1996).
\bibitem{quevedo} Hernando Quevedo, J.Math.Phys. {\bf 48}:  013506 (2007); Gen.Rel.Grav. {\bf 40}:971-984 (2008); arXiv:1111.5056 [math-ph].
\bibitem{liu} Haishan Liu, H. Lu, Mingxing Luo and Kai-Nan Shao, JHEP {\bf 1012}: 054 (2010); arXiv:1008.4482 [hep-th].
\bibitem{cai} Rong-Gen Cai and Yuan-Zhong Zhang, Phys.Rev.D {\bf 54}: 4891-4898 (1996); arXiv:gr-qc/9609065v1.
\bibitem{lemos} Jos\'{e} P. S. Lemos and Francisco S. N. Lobo, Phys.Rev. D {\bf 69}: 104007 (2004); arXiv:gr-qc/0402099v2.
\bibitem{cai2} Rong-Gen Cai, Jeong-Young Ji and Kwang-Sup Soh, Phys.Rev.D {\bf 57}:6547-6550 (1998); arXiv:gr-qc/9708063v2. 
\bibitem{gerard3} G\'{e}rard Cl\'{e}ment, J\'{u}lio C. Fabris and Manuel E. Rodrigues,  Phys. Rev. D {\bf 79}, 064021 (2009); arXiv:hep-th/09014543.
\bibitem{hannestad} S. Hannestad, Int. J. Mod. Phys. A {\bf 21}, 1938 (2006); J. Dunkley et al., Astrophys. J. Suppl. Ser. {\bf 180}, 306 (2009).
\bibitem{kirill1} K.A. Bronnikov, M.S. Chernakova, J.C. Fabris, N. Pinto-Neto and M.E. Rodrigues, Int.J.Mod.Phys.D {\bf 17}:25-42 (2008).
\bibitem{gibbons} G.W. Gibbons and D. A. Rasheed, Nucl. Phys. B {\bf 476}, 515 (1996).
\bibitem{gao} C. J. Gao and S. N. Zhang, arXiv:hep-th/0604114.
\bibitem{grojean} C. Grojean, F. Quevedo, G. Tasinato, and I. Zavala, J. High Energy Phys. {\bf 08}: 005 (2001).
\bibitem{hull} C. M. Hull, JHEP {\bf 9807}: 021 (1998).
\bibitem{gerard2} Mustapha Azreg-Ainou, G\'{e}rard Cl\'{e}ment, J\'{u}lio C. Fabris and Manuel E. Rodrigues, Phys.Rev.D {\bf 83}:124001 (2011).
\bibitem{davies2} N. D. Birrell and P. C. W. Davies, Quantum fields in curved space, Cambridge University Press (1982).
\bibitem{ford} L. H. Ford, in Particles and Fields: Proceedings of the
IXth Jorge Andre Swieca Summer School, Brazil, 16-28 February 1997, edited by J. C. Barata, Sergio F. Novaes and Adolfo P. C. Malbouisson (World Scientific, Singapore, 1998), pp. 345-388; arXiv: gr-qc/9707062.
\bibitem{hawking2} G. W. Gibbons and S. Hawking, Phys. Rev. D {\bf 15}: 2752-2756 (1977).
\bibitem{gerard4} G. Cl\'{e}ment, J. C. Fabris and G. T. Marques, Phys. Lett. B {\bf 651}: 54-57 (2007).
\bibitem{kanti} Panagiota Kanti and John March-Russell, Phys.Rev.D {\bf 66}: 024023 (2002); Wontae Kim and John J. Oh, J.Korean Phys.Soc. {\bf 52}: 986 (2008); Kazuo Ghoroku and Arne L. Larsen, Phys.Lett. B {\bf 328}: 28-35 (1994).
\bibitem{robinson} S.P. Robinson and F. Wilczek, Phys. Rev. Lett. {\bf 95},011303 (2005).
\bibitem{jacobson} T. Jacobson and G. Kang, Class.Quant.Grav. {\bf 10}:L201-L206 (1993); arXiv: gr-qc/9307002.
\bibitem{glauber} Glauber T. Marques and Manuel E. Rodrigues, Eur.Phys.J. C {\bf 72}:  1891 (2012); arXiv:1110.0079 [gr-qc]. 
\bibitem{hermann} R. Hermann, Geometry, physics and systems (Marcel Dekker, New York, 1973); G. Hern´andez and E. A. Lacomba, Contact Riemannian geometry and thermodynamics, Diff. Geom. and Appl. {\bf 8}, 205 (1998).
\bibitem{zui} Manuel E. Rodrigues and Zui A. A. Oporto, Phys. Rev. D {\bf 85}: 104022 (2012); arXiv:1201.5337v3 [gr-qc]
\bibitem{visser} Matt Visser, Lorentzian Wormholes, American Institute of Physics Press, 1995, New
York.
\bibitem{babichev} E. Babichev, V. Dokuchaev and Yu. Eroshenko, Phys.Rev.Lett. {\bf 93}:  021102 (2004); arXiv:gr-qc/0402089.

\bibitem{bronnikovPRL}K.A. Bronnikov and J.C. Fabris, Phys.Rev.Lett. {\bf 96}: 251101 (2006); arXiv:gr-qc/0511109.

\bibitem{phantom}Songbai Chen, Jiliang Jing, JHEP {\bf 0903}:081 (2009); K. A. Bronnikov, R. A. Konoplya, A. Zhidenko, Phys. Rev. D {\bf 86}: 024028 (2012); S.V. Bolokhov, K.A. Bronnikov, M.V. Skvortsova, Class.Quant.Grav. {\bf 29}: 245006 (2012); Mustapha Azreg-Ainou, Phys. Rev. D {\bf 87}: 024012 (2013); Galin N. Gyulchev, Ivan Zh. Stefanov, Phys. Rev. D {\bf 87}: 063005 (2013); Anna Nakonieczna, Marek Rogatko, Rafal Moderski, Phys.Rev.D {\bf 86}: 044043 (2012); Songbai Chen, Jiliang Jing, arXiv:1301.1440 [gr-qc].


\end{thebibliography}
\end{document}